\documentclass{article}
\usepackage{dcase2024,amsmath,graphicx,url,times,booktabs, tabularx,dsfont,graphicx,hyperref}
\usepackage{float}
\usepackage{xcolor}
\usepackage{pifont}
\usepackage{multirow}

\newcommand{\norm}[1]{\left\lVert#1\right\rVert}
\newcommand{\cmark}{\ding{51}}%
\newcommand{\xmark}{\ding{55}}%

\title{Estimated Audio--Caption Correspondences \\improve Language-Based Audio Retrieval}

\name{Paul Primus$^1$, Florian Schmid$^1$, and Gerhard Widmer$^{1,2}$}
\address{
$^1$Institute of Computational Perception (CP-JKU)  \\
$^2$LIT Artificial Intelligence Lab\\
Johannes Kepler University, Austria
}

\begin{document}

\ninept
\maketitle

\begin{sloppy}

\begin{abstract}
Dual-encoder-based audio retrieval systems are commonly optimized with contrastive learning on a set of matching and mismatching audio--caption pairs. This leads to a shared embedding space in which corresponding items from the two modalities end up close together. Since audio--caption datasets typically only contain matching pairs of recordings and descriptions, it has become common practice to create mismatching pairs by pairing the audio with a caption randomly drawn from the dataset. This is not ideal because the randomly sampled caption could, just by chance, partly or entirely describe the audio recording. However, correspondence information for all possible pairs is costly to annotate and thus typically unavailable; we, therefore, suggest substituting it with \textit{estimated correspondences}. To this end, we propose a two-staged training procedure in which multiple retrieval models are first trained as usual, i.e., without estimated correspondences. In the second stage, the audio--caption correspondences predicted by these models then serve as prediction targets. We evaluate our method on the ClothoV2 and the AudioCaps benchmark and show that it improves retrieval performance, even in a restricting self-distillation setting where a single model generates and then learns from the estimated correspondences. We further show that our method outperforms the current state of the art by 1.6 pp. mAP@10 on the ClothoV2 benchmark. 

\end{abstract}

\section{Introduction}
Language-based audio retrieval systems search for audio recordings based on textual descriptions. 
Such systems are of practical interest because they allow users to intuitively specify arbitrary acoustic concepts of interest (such as acoustic events, qualities of sound, and temporal relationships) without relying on a predefined set of tags or categories.
However, language-based retrieval is difficult from a technical perspective because it requires deriving comparable semantic representations for raw audio signals and sequences of words.
Typical audio retrieval systems \cite{journal,pooling_and_objective,loss_comparison,recent_work_2} achieve this via a dual-encoder architecture that projects the textual query and the candidate audio recordings into a shared multi-modal metric space where the audio recordings are then ranked based on their distance to the textual query (for a different approach, see previous work by Labb\'{e} et al. \cite{killing_birds}). 

\begin{figure}[ht]
    \centering
    \def\svgwidth{0.95 \linewidth}
    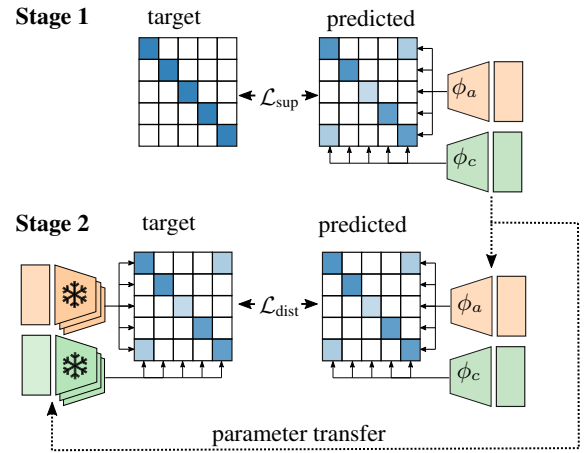
    \caption{Audio and descriptions are transformed into the shared audio--caption embedding space via the audio and description embedding models $\phi_\mathrm{a}$ and $\phi_\mathrm{c}$, respectively. In stage 1, we assume that audio $a_i$ and caption $c_j$ do not match if $i \neq j$ and train the model with contrastive loss $\mathcal{L}_{\textrm{sup}}$. Stage 2 uses predictions ensembled from several Stage 1 models (bottom left) to estimate the correspondence between $a_i$ and $c_j$; those estimates then serve as prediction targets instead of the ground truth from stage 1. Stage 2 model parameters are initialized with stage 1 parameters, and the corresponding loss is denoted as $\mathcal{L}_{\mathrm{dist}}$.}
    \label{fig:overview}
\end{figure}

Previous studies have explored multiple directions to improve language-based audio retrieval systems, such as using better pre-trained embedding models \cite{DCASE2023}, augmentation techniques for both audio and text \cite{DCASE2022}, artificial captions generated with large language models \cite{wavcaps,cacophony,DCASE2023}, or hybrid content and metadata based retrieval systems \cite{primus_eusipco2024}. 
In this work, we expand on the previously proposed idea of utilizing non-binary audio--caption correspondences for training retrieval models \cite{Xie2023_relevances}. However, instead of relying on crowd-sourced correspondences, our method estimates them via an ensemble of audio retrieval models. To this end, we propose a two-step training procedure that is illustrated in Figure \ref{fig:overview}. In the following sections, we motivate and describe the proposed two-stage training strategy; we then detail the experimental setup and present results on ClothoV2 \cite{clotho} and AudioCaps \cite{audiocaps}. When trained with large audio--caption datasets, our method outperforms the current state of the art on ClothoV2 by around $1.6$ pp. mAP@10.
Our submission to the DCASE Challenge 2024 \cite{primus2024_t8}, based on the proposed method, took the first rank in task 8.
Our implementation, model checkpoints, predictions, and examples are available on GitHub\footnote{\href{https://github.com/OptimusPrimus/salsa}{https://github.com/OptimusPrimus/salsa}}.

\section{Text-based Audio Retrieval}

Language-based retrieval systems typically consist of two modality encoder networks, one for audio and one for caption embedding, denoted as $\phi_a(\cdot)$ and $\phi_c(\cdot)$, respectively.
These encoders learn to embed recordings and descriptions into a shared $D$-dimensional embedding space such that representations of matching audio snippets and captions are similar. 
The agreement between audio $a_i$ and description $c_j$ at training or inference time is estimated via the normalized dot product in the multi-modal embedding space:
$$C_{ij} = \frac{\phi_{\textrm{a}}(a_{i})^T \cdot \phi_{\textrm{c}}(c_{j})}{\norm{\phi_{\textrm{a}}(a_{i})}^2 \norm{\phi_{\textrm{c}}(c_{j})}^2}$$
Previous research typically relied on contrastive learning to train audio retrieval models. A usual choice is an adapted version of the Normalized Temperature-scaled cross-entropy (NT-Xnt) loss \cite{NTxent}, which converts those agreements into conditional probability distributions over audio snippets and captions via a temperature-scaled softmax operation, where
$$q_a(a_i \mid c_j) = \frac{e^{C_{ij}/\tau}}{\sum_{i=1}^{N}e^{C_{ij}/\tau}}$$
gives the estimated probability that audio $a_i$ corresponds to a given caption $c_j$, and
$$q_c(c_j \mid a_i) = \frac{e^{C_{ij}/\tau}}{\sum_{j=1}^{N}e^{C_{ij}/\tau}}$$
gives the estimated probability that caption $c_j$ corresponds to a given audio $a_i$.
The training objective is then to minimize the cross-entropy (denoted as $H$) between the estimated and the actual correspondence probabilities, $q$ and $p$, respectively. 
$$\mathcal{L_\textrm{sup}} = H(p_a, q_{a}) + H(p_c, q_{c})$$
However, the true correspondence probabilities $p$ for audio $a_i$ and caption $c_j$ with $i \neq j$ are not generally available because audio retrieval datasets (e.g., \cite{clotho, audiocaps,wavcaps}) typically only provide a set of $N$ matching audio and caption pairs  $\{(a_i, c_i)\}_{i=1}^{N}$, but no correspondence annotations for the case $i \neq j$.  
Previous studies thus assumed that $c_j$ does not describe $a_i$ if $i \neq j$, which is reasonable if the dataset holds a large variety of recordings with very specific descriptions.
Using this assumption, the target probability distributions $p$ for recordings and captions 
can then be defined as follows:
$$p_a(a_i \mid c_j) := \mathds{1}_{i=j} \;\textrm{and}\; p_c(c_j \mid a_i) := \mathds{1}_{i=j}$$
Similar to Xie \cite{Xie2023_relevances}, we argue that relying on this assumption is not ideal, mainly for two reasons: \begin{enumerate}
    \item It is only valid if each caption in the dataset describes \textit{exactly one} recording, which is not the case for popular audio retrieval datasets such as ClothoV2, AudioCaps, and WavCaps, as demonstrated in \cite{primus2024_t8}.
    \item Binary correspondences are limited to modeling exact matches between audio recordings and captions. However, we believe that incentivizing the model to place partially matching captions closer to the corresponding audio recording in the multi-modal embedding space is beneficial.
\end{enumerate} 

Xie et al. \cite{Xie2023_relevances} crowdsourced pairwise correspondence scores of audios snippets and captions in a previous study but did not find significant benefits when incorporating binarized versions of those scores during training. We still hypothesize that additional correspondence annotations can provide useful guidance during training; however, there are no large-scale datasets with complete correspondence annotations due to the high cost associated with annotating $N^2$ audio--caption pairs for large $N$.

\section{Proposed Method}
\label{sec:method}

To obtain audio--caption correspondences without relying on human annotators, we suggest estimating them with an ensemble of $M$ independently pre-trained audio retrieval models. We chose to train those models as described in the previous section; however, other approaches like the method proposed in \cite{killing_birds} might lead to comparable results. 
In our setup, the predicted pairwise agreements are ensembled as follows:
$$\hat{C}_{ij} = \frac{1}{M}\sum_{m=1}^{M}{C^m_{ij}}$$ 
We use a softmax operation to convert those agreements to an estimate of the true correspondence probabilities of recordings given a caption
$$\hat{p}_{a}(a_i \mid c_j) := \frac{e^{\hat{C}_{ij}/\tau}}{\sum_{i=1}^{N}e^{\hat{C}_{ij}/\tau}}$$
and an estimate of the true correspondence probabilities of captions given an audio
$$\hat{p}_{c}(c_j  \mid a_i) := \frac{e^{\hat{C}_{ij}/\tau}}{\sum_{j=1}^{N}e^{\hat{C}_{ij}/\tau}}$$
These two probability distributions then serve as prediction targets instead of the deterministic correspondence probabilities $p_a$ and $p_c$ in the NT-Xent loss. We refer to the corresponding loss as distillation loss
$$\mathcal{L}_{\textrm{dist}} = H(\hat{p}_a, q_{a}) + H(\hat{p}_c, q_c)$$
due to the conceptual similarity to knowledge distillation \cite{knowledge_distilation}.
\section{Experimental Setup}

\begin{table*}[ht]
\centering
\begin{tabular}{@{}lll|llll|llll@{}}
\toprule
audio     &             &  & \multicolumn{4}{c|}{ClothoV2}  & \multicolumn{4}{c}{AudioCaps}  \\
embedding &  $\hat{p}$  & $M$ & mAP@10 & R@1   & R@5   & R@10  & mAP@10 & R@1   & R@5   & R@10  \\ \midrule
PaSST     & \xmark & -      & 28.93  & 18.11 & 43.54 & 57.57 & 55.30  & 40.74 & 75.89 & 86.28 \\
PaSST     & \cmark & 3      & 31.25  & 19.52 & 46.49 & 61.30 & 57.61  & 42.55 & 79.04 & 88.74 \\
PaSST     & \cmark & 1      & 30.18  & 18.95 & 45.28 & 59.43 &   56.67     &   41.66    &  78.00     &  88.21     \\ \midrule
ATST      & \xmark & -      & 28.26  & 17.36 & 42.58 & 56.25 & 55.86  & 42.47 & 74.06 & 84.26 \\
ATST      & \cmark & 3      & 31.01  & 19.29 & 46.33 & 61.07 & 59.37  & 45.43 & 78.02 & 88.51 \\
ATST      & \cmark & 1      & 29.83  & 17.75 & 43.73 & 57.81 &  57.72      &   43.89    &  77.17    &   86.96    \\ \midrule
MN        & \xmark & -      & 28.72  & 17.57 & 43.73 & 57.82 & 55.16  & 40.26 & 74.06 & 87.11 \\
MN        & \cmark & 3      & 30.25  & 18.49 & 46.11 & 59.81 & 57.06  & 41.83 & 77.92 & 88.45 \\
MN        & \cmark & 1      & 28.96  & 17.80 & 44.59 & 57.84 &  54.95      &    39.62   &  76.66     &    88.32   \\ \midrule
hybrid MN~\cite{primus_eusipco2024}      & \xmark & -      & 29.88  & 18.39 & 45.04 & 58.62 & 58.61  & 43.47 & 79.38 & 90.16 \\ \bottomrule
\end{tabular}
\caption{Retrieval performance on the AudioCaps and Clotho benchmarks. Each section corresponds to a different Audio Embedding Model. Results in the first row in each section correspond to results without estimated audio--caption correspondences (i.e., \xmark\;in column $\hat{p}$ ). The second row gives results of models fine-tuned with the estimated audio--caption correspondences (i.e., \cmark\;in column $\hat{p}$ and $M=3$). The third row gives results in the self-distillation setting (i.e., \cmark in column $\hat{p}$ and $M=1$).}
\label{tab:main_results}
\end{table*}

\subsection{Datasets \& Benchmarks}

We experimented with two popular audio-retrieval benchmark datasets, namely ClothoV2 \cite{clotho} and AudioCaps \cite{audiocaps}. We additionally use WavCaps \cite{wavcaps} for training to compare our method to the current state of the art. We briefly describe the datasets below.

ClothoV2 \cite{clotho} contains $15$-$30$ second recordings and captions that are between 8 and 20 words long. The provided training, validation, and test split contain 3840, 1045, and 1045 recordings, respectively; each recording is associated with five human-generated captions.

AudioCaps \cite{audiocaps} consists of $51,308$ audio recordings taken from AudioSet \cite{audioset}. Each training and validation recording is associated with one and five human-written captions, respectively. The audio recordings' length is roughly 10 seconds, and the captions are, on average, 9.8 words long.

WavCaps \cite{wavcaps} is currently the largest audio--caption dataset available; it contains $403,050$ audio recordings and has been used successfully in previous studies to reach state-of-the-art retrieval performance \cite{wavcaps, VAST, DCASE2023}.
Each audio file in WavCaps is associated with a synthetic audio caption that was created by instructing the GPT3.5-turbo model to extract relevant sound events from metadata and output a single-sentence description. The generated captions are, on average, 7.8 words long.  In order to avoid information leakage between the training and evaluation sets, we excluded the overlapping recordings between WavCaps and the evaluation sub-sets of ClothoV2.

\subsection{Pretrained Embedding Models}
We experimented with three audio embedding models (PaSST \cite{passt}, ATST \cite{li2022atstf}, and MN \cite{mn40_as}) and one text embedding model (RoBERTa \cite{roberta}); below, we briefly describe how we used them for audio and text embedding. 

\subsubsection{Audio Embedding}
PaSST \cite{passt} has a positional encoding for inputs of up to $10$ seconds; we thus cut the up to 30-second long inputs into non-overlapping 10-second snippets and averaged their embeddings. We used the version of PaSST without patch overlap and applied structured patchout of 2 and 15 patches over frequency and time dimensions, respectively. 
We used the checkpoint denoted as \texttt{passt\_s\_p16\_s16\_128\_ap468} in our experiments, which is available via GitHub\footnote{\href{https://github.com/kkoutini/PaSST}{https://github.com/kkoutini/PaSST}}.

ATST-Frame \cite{li2022atstf} (denoted only as ATST in the following) has a positional encoding that is also limited to $10$ seconds; we again cut the audio recordings into non-overlapping 10-second snippets and averaged their embeddings to obtain a single embedding vector. During training, we used frequency warping \cite{li2022atstf} where at most 10\% of the higher frequency bins were dropped. We used a publicly available checkpoint of ATST (called \texttt{atst\_as2M.ckpt}) that was further fine-tuned on the weak labels of AudioSet\footnote{\href{https://github.com/Audio-WestlakeU/ATST-SED}{https://github.com/Audio-WestlakeU/ATST-SED}}.

EfficientAT MobileNetV3 \cite{mn40_as} (referred to as MN in the following) is particularly well suited for experiments with ClothoV2 because the CNN architecture can handle audio recordings of arbitrary length as input. We used the model with ID \texttt{mn40\_as\_ext} in our experiments. The checkpoint is available on GitHub\footnote{\href{https://github.com/fschmid56/EfficientAT}{https://github.com/fschmid56/EfficientAT}}. 

\subsubsection{Sentence Embedding}
Roberta large \cite{roberta} was used for sentence embedding because it gave the best performance in our previous comparison of text embedding models~\cite{DCASE2023}. RoBERTa is a bi-directional self-attention-based sentence encoder that underwent self-supervised pretraining on the BookCorpus \cite{bookcorpus} and WikiText datasets \cite{wikitext}.
The RoBERTa large model has around 354 million parameters. 

\subsection{Optimization}
During pre-training (stage 1), both modality encoders were jointly optimized using gradient descent with a batch size of 64 for PaSST and ATST and 32 for MN. We used the Adam update rule \cite{adam} to minimize $\mathcal{L}_{\textrm{sup}}$ for 20 epochs, with one warmup epoch. Thereafter, the learning rate was decayed from $2 \times 10^{-5}$ to $10^{-7}$ using a cosine schedule. The hyperparameters of the optimizer were set to PyTorch's \cite{pytorch} defaults.

Fine-tuning (stage 2) was done by minimizing $\mathcal{L}_{\textrm{dist}}$. Model parameters in stage 2 were initialized with the parameters from stage 1. The training schedule and learning rate were chosen to be the same as in Stage 1 (however, they might benefit from additional tuning). Audio--caption correspondence estimates were obtained by assembling the similarity scores of all three models ($M=3$) as described in Section \ref{sec:method}. We set $\tau$ to a constant value of $0.05$ in all our experiments. 

We used the benchmarks' validation sets to select checkpoints and report results on the test sets here. Our main evaluation criterion for hyperparameter selection was the mean Average Precision among the top-10 results (mAP@10) which is the metric used for ranking systems in the DCASE Challenge. In the results section, we additionally report the recall among the top-1, top-5, and top-10 retrieved results, which allows more detailed analysis and comparison with additional previous work.

\section{Results \& Discussion}
\label{section:results}

Table~\ref{tab:main_results} summarizes the retrieval performance of our method on the AudioCaps and ClothoV2 benchmarks. Each section in Table~\ref{tab:main_results} corresponds to one of the three audio embedding models. We chose to experiment without external data first to demonstrate the effectiveness of our method. In Section~\ref{sec:scaling}, we will then show that our method establishes new state-of-the-art performance on the ClothoV2 benchmark when paired with a large audio--caption dataset.
 \vspace{-4pt}
\subsection{Does fine-tuning with estimated correspondences lead to improved retrieval performance?}
 \vspace{-4pt}
We first pre-trained the three retrieval models without estimated correspondence and report the results in the first row of each section of Table\;\ref{tab:main_results}. The resulting models were then fine-tuned using the ensembled audio--caption correspondence estimates of all three retrieval models from stage 1. The results are given in the second row of each section in Table\;\ref{tab:main_results}. We note a substantial increase across all performance metrics for both ClothoV2 and AudioCaps, which indicates that using estimated correspondences has a positive effect. 

We additionally compare the proposed method to our recent hybrid content and metadata-based retrieval system \cite{primus_eusipco2024}, denoted as hybrid MN in Table\;\ref{tab:main_results} (last section). We find that using the estimated correspondences leads to similar improvements on both benchmarks, but without relying on additional audio metadata such as descriptive tags for retrieval. We hypothesize that combining these two approaches could lead to further performance gains. 

\subsection{Ablation Study: Is a diverse ensemble required to achieve improvements with estimated audio--caption correspondences?}

In the previously described experiments, we relied on ensembled predictions from three diverse models ($M=3$) to derive the audio--caption correspondences. 
We want to understand if the performance improvement is a result of distilling from an ensemble of multiple models or if similar results can be achieved in a self-distillation setting. To this end, we dropped the ensembling of multiple models when deriving the correspondences, i.e., we used the same model to generate and then learn from the estimated correspondences.
The results are given in the third row of each section in Table~\ref{tab:main_results}. We observe that PaSST and ATST benefitted even in this limiting self-distillation setting. However, we also note that MN's performance did not generally improve over the pretraining performance. We hypothesize that this could be fixed with the additional hyperparameter tuning for the second stage.
We further observe that using ensemble predictions led to an additional performance improvement over the self-distillation approach (compare rows two and three in each section). We thus recommend using ensembled predictions to estimate audio--caption correspondences whenever additional models are available.

\subsection{Comparison to state-of-the-art systems}
\label{sec:scaling}

Current state-of-the-art audio retrieval systems \cite{DCASE2023, VAST} train on multiple audio--caption datasets to increase their performance. To compare our method to these systems under fair conditions, we also increased the size of the training set. To this end, we combined AudioCaps, ClothoV2, and WavCaps (as done in \cite{DCASE2022}) and pretrained the three previously introduced systems on the merged dataset. The resulting models were fine-tuned on ClothoV2 by minimizing a linear combination of $\mathcal{L}_\textrm{sup}$ and $\mathcal{L}_\textrm{dist}$. We conducted a grid search over the linear combination's weight, the learning rate, and possible ensemble combinations and selected the best PaSST model on the ClothoV2 validation set.

The first section in Table~\ref{tab:comparison} compares the performance of models before and after fine-tuning on ClothoV2. Stage 1 training on the scaled-up dataset (first row in Table\;\ref{tab:comparison}) already led to better results than training only on ClothoV2. When this model was fine-tuned on ClothoV2 without the estimated correspondences (second row in Table\;\ref{tab:comparison}), the mAP@10 improved by around 0.9 pp; when the estimated correspondences were used during fine-tuning (third row in Table\;\ref{tab:comparison}), the mAP@10 increased even more, namely by around $4.6$ pp.

The second section in Table \ref{tab:comparison} compares our method to current state-of-the-art audio retrieval systems. Our proposed method outperforms last year's best single system submission to the DCASE Challenge (Submission 2 of \cite{primus2023_t6b}) by around $1.6$ pp. without using text augmentations and synthetic captions.  The results also show that our approach achieves a higher recall compared to VAST \cite{VAST}, a vision--audio--text model that was trained on 27 million videos.

\begin{table}[ht]
\centering
\begin{tabular}{@{}ll|llll@{}}
\toprule
                            &  & \multicolumn{4}{c}{ClothoV2}   \\
method                      & $\hat{p}$ & mAP@10 & R@1   & R@5   & R@10  \\ \midrule
PaSST (stage 1)             & \xmark & 35.46      & 23.64  & 51.44  & 64.98  \\
PaSST (stage 2)             & \xmark & 36.33      & 24.31  & 52.84  & 65.63  \\
PaSST (stage 2)             & \cmark & \textbf{40.14}  & \textbf{27.69} & \textbf{57.03} & \textbf{70.39} \\ \midrule
DCASE23 \cite{primus2023_t6b} & \xmark & 38.56  & 26.07 & 55.27 & 69.30 \\
VAST \cite{VAST}            & - & - & 26.9  & 53.2  & 66.1  \\ \bottomrule
                   
\end{tabular}
\caption{First section: Performance of our method on the ClothoV2 benchmark when models were pre-trained on WavCaps, AudioCaps, and ClothoV2. A \cmark\;in column $\hat{p}$ indicates that estimated correspondences were used when fine-tuning on ClothoV2 in stage 2. Second section: Performance of current state-of-the-art audio-retrieval models. }
\label{tab:comparison}
\end{table}


\section{Conclusion}
In this work, we have explored the use of estimated audio--caption correspondences to train language-based audio retrieval models. We proposed a two-stage training procedure that first estimates the correspondences and then uses those estimated correspondences for training. We showed that ensemble correspondence estimates lead to improved retrieval performance on both AudioCaps and ClothoV2. We further experimented with using the same model to generate
and then learn from the estimated correspondences, which led to improved performance for two out of the three investigated retrieval systems.
Finally, we scaled up our approach by combining multiple datasets; the resulting model outperforms the previous state-of-the-art on ClothoV2 by around 1.6 pp. mAP@10.

\section{ACKNOWLEDGMENT}
\label{sec:ack}
The LIT AI Lab is financed by the Federal State of Upper Austria. Gerhard Widmer's work is supported by the European Research Council (ERC) under the European Union's Horizon 2020 research and innovation programme, grant agreement No 101019375 (Whither Music?). Some parts of the computational results presented in this work have been achieved using the Vienna Scientific Cluster (VSC).

\bibliographystyle{IEEEtran}
\bibliography{refs}

\end{sloppy}
\end{document}